# Set-based MPC for discrete-time LTI systems with maximal domain of attraction and minimal predictive control horizon

Alejandro Anderson[a], Agustina D'Jorge[a], Alejandro H. González[a], Antonio Ferramosca[b], Marcelo Actis[c]

[a] Institute of Technological Development for the Chemical Industry (INTEC), CONICET-Universidad Nacional del Litoral (UNL). Güemes 3450, (3000) Santa Fe, Argentina.
[b] Department of Management, Information and Production Engineering, University of Bergamo. Via Marconi 5, 24044 Dalmine (BG), Italy.
[c] Facultad de Ingeniería Química, UNL-CONICET, Santiago del Estero 2829, (3000) Santa Fe, Argentina.

**Corresponding Author:**

Marcelo Actis
Facultad de Ingeniería Química, UNL–CONICET
—
Postal Address:
FIQ, UNL, Santa Fe
Santiago del Estero 2829.
S3000AOM, Santa Fe, ARGENTINA.
—
Phone: +54 342 4580 3164
e-mail: mactis@fiq.unl.edu.ar

# Set-based MPC for discrete-time LTI systems with maximal domain of attraction and minimal predictive control horizon

## Abstract

This paper presents a novel set-based model predictive control for tracking, which provides the largest domain of attraction, even with the minimal predictive/control horizon. The formulation - which consists of a single optimization problem - shows a dual behavior: one operating inside the maximal controllable set to the feasible equilibrium set, and the other operating at the $N$-controllable set to the same equilibrium set. Based on some finite-time convergence results, asymptotic stability of the resulting closed-loop is proved, while recursive feasibility is ensured for any change of the setpoint. The properties and advantages of the proposal have been tested on simulation models.

## 1. Introduction

Model Predictive Control (MPC) is a strategy widely used in industries since the 1980s, due to its ability to deal with multivariable processes including both, state and input constraints. A strong theoretical framework has been developed in the last two decades, showing that MPC is a control technique capable to guarantee asymptotic stability, constraint satisfaction and robustness, based on the solution of an on-line tractable optimization problem for both, linear and nonlinear systems. Lyapunov theory provides a general framework to prove asymptotic stability of a system controlled by an MPC, by employing suitable terminal ingredients (terminal cost, terminal equality or inequality constraint) ([10, 23, 22]).

The stabilizing terminal constraint implicitly imposes hard restrictions on the state, since only those states that can be steered in a given number of steps to the terminal region will be properly stabilized. These states determine the so-called closed-loop domain of attraction, whose characterization is crucial because it represents a domain of validity that is determined not just by the system dynamic and its constraints, but also by the specific controller design. In this context, any effort to modify the classical MPC formulation to have a larger domain of attraction is remarkably beneficial, as it was stated in many seminal works ([16, 19]).

The most obvious way to enlarge the MPC domain of attraction is by enlarging the prediction/control horizon. However, this strategy has two major drawbacks. Firstly, it produces a significant increase of the computational burden. Secondly, it is not unusual that during the operation of the plant the desired setpoint where the system has to be stabilized may change. As a result, the stabilizing terminal ingredients may become invalid, the feasibility of the MPC optimization problem may be lost, and the controller may fail to track the desired setpoint.

To solve the loss of feasibility under setpoint changes, different solutions have being proposed in the literature, such as *reference governors MPC* ([25, 5]) and *MPC for Tracking (MPCT)* ([20, 8, 18, 21]). This latter strategy solves the problem of recursive feasibility by penalizing the distance from the predicted trajectories to some extra artificial optimization variables, which are forced to be a feasible equilibrium. This way, not only the recursive feasibility is ensured for any change of setpoints, but also the domain of attraction is enlarged (although it does not necessarily reach the maximal domain of attraction, for a given prediction horizon), since it results to be the N-step controllable set to any equilibrium point of the system to be controlled, and not just the a specific one.

Another approach, tending to enlarge the MPC domain of attraction in a rather theoretical form, was presented in [19]. The idea was to substitutes the terminal constraint by a kind of contractive terminal constraint, which forces the terminal state to pass from one control invariant set to another. The proposed method reaches the maximal domain of attraction (the so-called *maximal controllable set to the equilibrium*, which is determined by the system and the constraints), but the computation of the control invariant sets, which may be computationally prohibitive, must be carried out on line every time a change in the setpoint occurs.

In this work we propose a novel MPC design based on set-based predictive control. In contrast to regular MPC in the so-called set dynamical systems framework ([6, 24]), the state of a system is identified by a set rather than a point. The set-based predictive controller generalizes conventional MPC along the lines of the tube based MPC approach [4, 9]. In particular, [4] is devoted to obtain provably safe controllers for disturbed nonlinear systems with constraints on states and inputs, while in [9] a reachable-set based version of the robust MPC based on tightened constraints originally proposed in [7] is presented. Some approaches use reachable sets based on zonotopes [12] to compute the reachable sets. However, it is worth to mention that the referred set-based strategies are not in the same control framework as the proposal. The aim of this work is to extend the domain of attraction up to the maximal controllable set for any fixed prediction horizon – even the minimum possible horizon – without loss of feasibility under changes of the setpoint. The method consists in a decomposition of the maximal domain of attraction into a disjoint union of embedded layers defined by the controllable sets of the system. The proposed controller shows a dual behavior, but into an unified formulation. First, the goal of the controller is to drive the system through the layers until a proper neighborhood of the equilibrium set is reached. This neighborhood is actually the domain of attraction of the classical MPCT [20]. Once the state of the system is inside this set, analogously to the classical MPCT, the asymptotic stability of (any point in) the equilibrium set is guaranteed.

This paper is organized as follows. We set up our notation in Section 1.1. In Section 2 we present general definitions and necessaries results to formulate the proposed MPC. For the sake of completeness we include in Section 2.1 a brief recall of the MPCT. Section 3 is devoted to describe in detail the proposed controller. The proof of the main results are addressed in Section 3.1. Finally, numerical simulations and conclusions can be found in Section 4 and 5, respectively.

### 1.1. Notation

We denote with $\mathbb{N}$ the sets of integers, $\mathbb{N}_0 := \mathbb{N} \cup \{0\}$ and $I_i := \{0, 1, \ldots, i\}$. The euclidean distance between two points $x, y \in \mathbb{R}^d$ by $\|x - y\| := [(x - y)'(x - y)]^{1/2}$, where $x'$ denotes the transpose of $x$. If $P$ is a positive-definite matrix on $\mathbb{R}^{d\times d}$ then we define the quadratic form $\|x - y\|_P^2 := (x - y)'P(x - y)$.

Let $\mathcal{X} \subseteq \mathbb{R}^d$. The *open ball with center in* $x \in \mathcal{X}$ *and radius* $\varepsilon > 0$ *relative to* $\mathcal{X}$ is given by $\mathcal{B}_{\mathcal{X}}(x, \varepsilon) := \{y \in \mathcal{X} : \|x - y\| < \varepsilon\}$. Given $x \in \Omega \subseteq \mathcal{X}$, we say that $x$ is an *interior point of* $\Omega$ *relative to* $\mathcal{X}$ if the there exist $\varepsilon > 0$ such that the open ball $\mathcal{B}_{\mathcal{X}}(x, \varepsilon) \subseteq \Omega$. The *interior of* $\Omega$ *relative to* $\mathcal{X}$ is the set of all interior points and it is denoted by $\operatorname{int}_{\mathcal{X}} \Omega$. In case $\mathcal{X} = \mathbb{R}^d$ we omit the subscript in the latter definition, i.e. $\operatorname{int} \Omega := \operatorname{int}_{\mathbb{R}^d} \Omega$. Finally, let $\Omega_1$ and $\Omega_2$ two sets in $\mathbb{R}^d$, we denote the difference between $\Omega_1$ and $\Omega_2$ by $\Omega_1 \setminus \Omega_2 := \{x \in \Omega_1 : x \notin \Omega_2\}$.

## 2. Problem statement and preliminaries

Consider a system described by a linear discrete-time invariant model

$$x(i + 1) = Ax(i) + Bu(i) \tag{2.1}$$

where $x(i) \in \mathcal{X} \subset \mathbb{R}^n$ is the system state at time $i$ and $u(i) \in \mathcal{U} \subset \mathbb{R}^m$ is the current control at time $i$, where $\mathcal{X}$ and $\mathcal{U}$ are compact convex sets containing the origin. As usual, the plant model is assumed to fulfill the following assumption:

**Assumption 2.1.** *The pair* $(A, B)$ *is controllable and the state is measured at each sampling time.*

The set of steady states and inputs of the system (2.1) is given by

$$\mathcal{Z}_s := \{(x_s, u_s) \in \mathcal{X} \times \mathcal{U} : \; x_s = Ax_s + Bu_s\}.$$

Thus, the equilibrium state and input sets are defined as

$$\mathcal{X}_s := \operatorname{proj}_{\mathcal{X}} \mathcal{Z}_s \quad \text{and} \quad \mathcal{U}_s := \operatorname{proj}_{\mathcal{U}} \mathcal{Z}_s.$$

The goal of this work is to propose a set based MPC which ensures at the same time maximal domain of attraction and feasibility/asymptotic stability under any change of setpoint. To this aim, let us first introduce the definitions of the main sets necessary for the formulation of the MPC proposed in Section 3.

**Definition 2.2 (Control Invariant Set).** *A set* $\Omega \subset \mathcal{X}$ *is a Control Invariant Set (CIS) of system* (2.1) *if for all* $x \in \Omega$*, there exists* $u \in \mathcal{U}$ *such that* $Ax + Bu \in \Omega$*. Associated to* $\Omega$ *is the corresponding input set*

$$\Psi(\Omega) := \{u \in \mathcal{U} : \exists\, x \in \Omega \text{ such that } Ax + Bu \in \Omega\}.$$

**Definition 2.3 ($i$-Step Controllable Set).** *Given $i \in \mathbb{N}$ and two sets $\Omega \subseteq \mathcal{X}$ and $\Psi \subseteq \mathcal{U}$, the i-step controllable set to $\Omega$ corresponding to the input set $\Psi$, of system* (2.1)*, is given by*

$$S_i(\Omega, \Psi) := \{x_0 \in \mathcal{X} : \forall\, j \in I_{i-1},\ \exists\, u_j \in \Psi \text{ such that } x_{j+1} = Ax_j + Bu_j \in \mathcal{X} \text{ and } x_i \in \Omega\}.$$

*For convenience, we define $S_0(\Omega, \Psi) := \Omega$ and $S_\infty(\Omega, \Psi) := \bigcup_{i=0}^{\infty} S_i(\Omega, \Psi)$, i.e. the set of admissible states which can be steered to the set $\Omega$ by a finite sequence of inputs in $\Psi \subseteq \mathcal{U}$.*

According to the latter definition, the maximal domain of attraction that the constrained system allows, for any setpoint $x^*$ in the equilibrium set $\mathcal{X}_s$, is given by $S_\infty(\mathcal{X}_s, \mathcal{U})$.

Next, we define a new special type of invariant sets, which are a key concept of this work.

**Definition 2.4 (Contractive CIS).** *Let $\Omega \subset S_\infty := S_\infty(\Omega, \mathcal{U})$ be a CIS. Then $\Omega$ is a contractive CIS if for all $x \in \Omega$, there exists $u \in \mathcal{U}$ such that $Ax + Bu \in \operatorname{int}_{S_\infty} \Omega$.*

Note that the above definition is similar to the definition of a $\gamma$-*Control Invariant Set*[1] (see [6]). Indeed, if $\Omega$ is a $\gamma$-control invariant set then for all $x \in \Omega$, there exists $u \in \mathcal{U}$ such that $Ax + Bu \in \operatorname{int} \Omega \subseteq \operatorname{int}_{S_\infty} \Omega$. Hence every $\gamma$-control invariant set is a contractive CIS. However the inverse result is not necessary true. For instance, it can be shown that the sets $S_{kN}$ corresponding to the double integrator presented in Section 4 are Contractive CIS but, due to the fact that they share some part of the boundary (see Figure 3), it cannot be ensured that they are $\gamma$-CIS. The importance of considering the weakened concept of interior relative to the set $S_\infty$ will be addressed in Remark 2.7.

**Lemma 2.5 (Geometric property of contractive CIS).** *Let $\Omega \subset S_\infty$ be a compact and convex contractive CIS of system* (2.1)*. Then, $\Omega \subseteq \operatorname{int}_{S_\infty} S_1(\Omega, \Psi(\Omega))$.*

*Proof.* It is easy to see that $\Omega \subseteq S_1(\Omega, \Psi(\Omega))$ by the invariance property of the set. It remains to show that every point of $\Omega$ is an interior point of $S_1(\Omega, \Psi(\Omega))$ relative to $S_\infty$. Let $x \in \Omega$, since $\Omega$ is a contractive CIS, there exists $u \in \Psi(\Omega)$ such that $Ax + Bu \in \operatorname{int}_{S_\infty} \Omega$. Then, there exists $\varepsilon > 0$ such that $\mathcal{B}_{S_\infty}(Ax + Bu, \varepsilon) \subseteq \Omega$. Since $Ax + Bu$ is a continuous function at $x$ from $S_\infty$ to $S_\infty$, then there exists $\delta > 0$ such that for all $\tilde{x} \in \mathcal{B}_{S_\infty}(x, \delta)$ we have

$$A\tilde{x} + Bu \in \mathcal{B}_{S_\infty}(Ax + Bu, \varepsilon) \subseteq \Omega.$$

Hence $\tilde{x} \in S_1(\Omega, \Psi(\Omega))$. Therefore $\mathcal{B}_{S_\infty}(x, \delta) \subseteq S_1(\Omega, \Psi(\Omega))$, that is to say so $\Omega \subseteq \operatorname{int}_{S_\infty}(S_1(\Omega, \Psi(\Omega)))$.

[1] $\Omega$ is a $\gamma$-Control Invariant Set ($\gamma$-CIS) if for $x \in \Omega$ there exists $u \in \mathcal{U}$ such that $Ax + Bu \in \gamma\Omega$, for some $\gamma < 1$.

The next lemma shows that the contractive invariance property is inheritable for the controllable sets.

**Lemma 2.6.** *Let $\Omega \subset \mathcal{X}$ be a compact and convex contractive CIS of system* (2.1)*. Then for every $i \in \mathbb{N}$, the set $S_i(\Omega, \mathcal{U})$ is a convex and compact contractive CIS of system* (2.1)*.*

*Proof.* Since $\Omega$ is under the assumptions of Lemma 2.5 and $\Psi(\Omega) \subseteq \mathcal{U}$ then

$$\Omega \subseteq \operatorname{int}_{S_\infty} S_1(\Omega, \Psi(\Omega)) \subseteq \operatorname{int}_{S_\infty} S_1(\Omega, \mathcal{U}).$$

Hence $S_1(\Omega, \mathcal{U})$ is a contractive CIS of system (2.1). By [15] we know that $S_1(\Omega, \mathcal{U})$ is compact and convex. Therefore $S_1(\Omega, \mathcal{U})$ is also under the assumptions of Lemma 2.5. The result follows by induction.

**Remark 2.7.** *Observe that in Lemma 2.5 we prove a geometric property of the contractive CIS analogous to Property 1 in [3] for $\gamma$-CIS. However in [3] it is required that $\Omega \subseteq \operatorname{int} \mathcal{X}$. This requirement represents an obstacle in the proof of Lemma 2.6 when we apply recursively Lemma 2.5 to the sets $S_i(\Omega, \mathcal{U})$. Notice that, for $i$ large enough, the sets $S_i(\Omega, \mathcal{U})$ usually collapse in the boundary of $\mathcal{X}$ (see Figure 3). Therefore they will not fulfill the hypothesis $S_i(\Omega, \mathcal{U}) \subseteq \operatorname{int} \mathcal{X}$.*

Next, we define a class of sets constituting a nested disjoint decomposition of the state space.

**Definition 2.8 ($k$-Layer Set).** *Let $\Omega \subset \mathcal{X}$ be a control invariant set, $\Psi \subseteq \mathcal{U}$ an input set and $N \in \mathbb{N}$. For any $k \in \mathbb{N}_0$ we define the $k$-Layer Set by $L_{kN}(\Omega, \Psi) := S_{(k+1)N}(\Omega, \Psi) \setminus S_{kN}(\Omega, \Psi)$.*

**Remark 2.9.** *In the above definition we ask for $\Omega$ to be an invariant set. This implies that the $i$-Step Controllable Sets $S_i(\Omega, \Psi)$, $i \in \mathbb{N}_0$, are nested (see Lemma 1 in [19]). Hence the $k$-Layer Sets are disjoint (see Figure 1) and even more*

$$S_\infty(\Omega, \Psi) = S_N(\Omega, \Psi) \cup \bigcup_{k=1}^{\infty} L_{kN}(\Omega, \Psi). \tag{2.2}$$

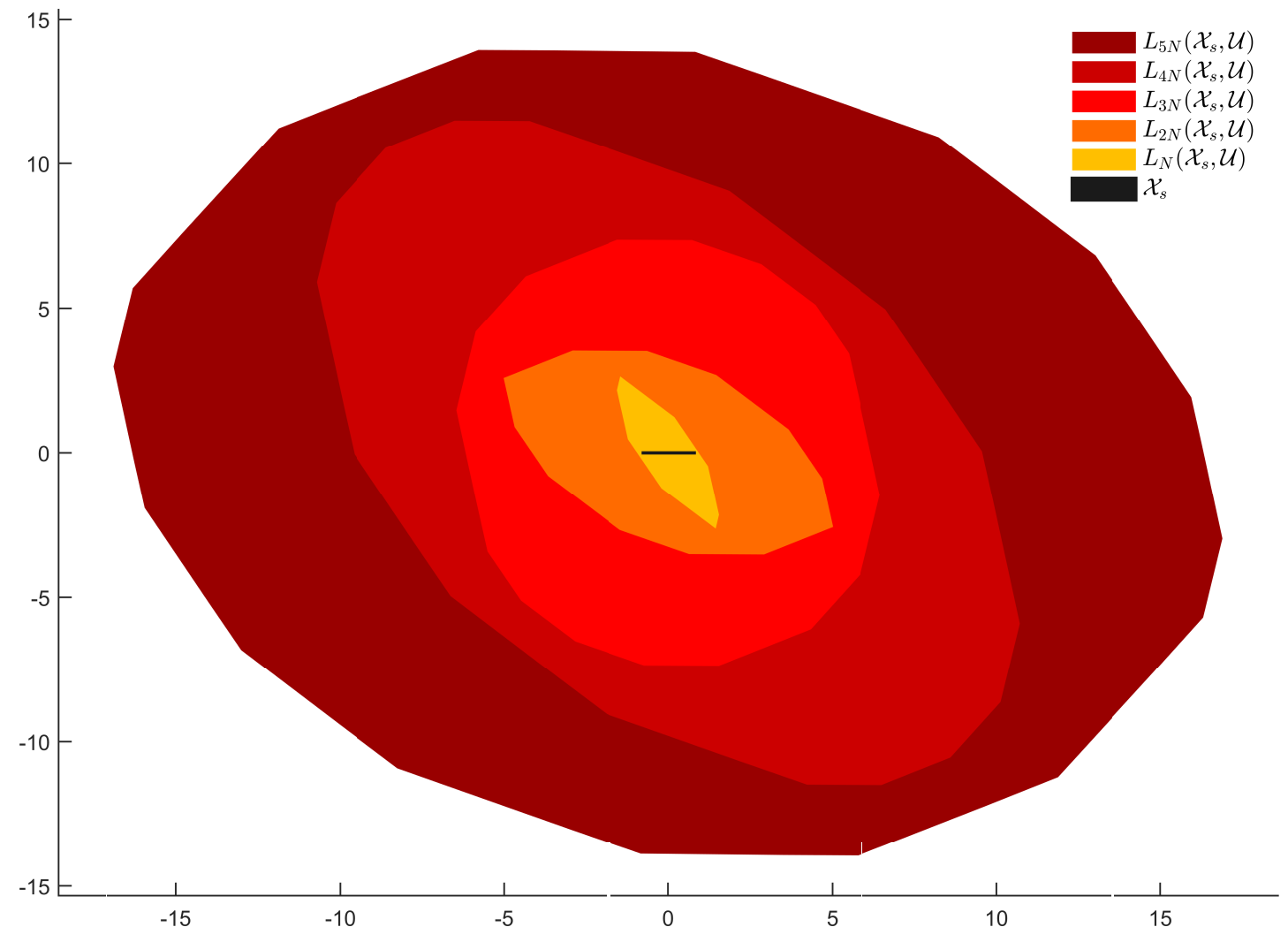

Figure 1: The first five layers for the harmonic oscillator system.

### *2.1. MPC for Tracking (MPCT)*

The MPC for tracking ([21]) attempts to track any admissible target steady sate by means of admissible evolutions, with an artificial reachable setpoint in $\mathcal{X}_s$ added as decision variable.

The MPCT cost function has two terms. The first one is a quadratic cost of the expected tracking error with respect to the artificial steady state and input $(x_s, u_s)$. The second one is the offset cost function $V_O(x_s, x^*)$, that penalizes the deviation from the artificial steady state $x_s$ to the desired setpoint $x^*$. The MPCT cost function is given by

$$V_N(x, x^*; \mathbf{u}, x_s, u_s) = \sum_{j=0}^{N-1} \|x_j - x_s\|_Q^2 + \|u_j - u_s\|_R^2 \\ + V_O(x_s, x^*), \qquad (2.3)$$

where $N$ is the control horizon, $\mathbf{u}$ is a sequence of $N$ future control inputs, i.e. $\mathbf{u} = \{u_0, \cdots, u_{N-1}\}$, $x_j$ is the predicted state of the system at time $j$ given by $x_{j+1} = Ax_j + Bu_j$, with $x_0 = x$. The offset cost function is required to be a convex and positive-definite function (see [21]). For simplicity, in this work we will considered it as a quadratic function $V_O(x_s, x^*) = \|x_s - x^*\|_T^2$, being $T$ a positive-definite matrix.

If a terminal equality constraint is used to ensure stability, the MPCT optimization problem $P_N^T(x, x^*)$ is given by

$$\min_{\mathbf{u}, x_s, u_s} V_N(x, x^*; \mathbf{u}, x_s, u_s) \qquad (2.4)$$

$$\text{s.t.} \quad x_0 = x, \tag{2.5}$$
$$x_{j+1} = Ax_j + Bu_j, \quad j \in I_{N-1}, \tag{2.6}$$
$$x_j \in \mathcal{X}, \quad u_j \in \mathcal{U}, \quad j \in I_{N-1}, \tag{2.7}$$
$$(x_s, u_s) \in \mathcal{Z}_s, \tag{2.8}$$
$$x_N = x_s. \tag{2.9}$$

Constraint (2.8) ensures that the artificial variable $(x_s, u_s)$ is an admissible equilibrium point. The terminal equality constraint (2.9) forces the terminal state to be the artificial state, that is to be the best (based on the offset cost function) admissible equilibrium point. These constraints ensure recursive feasibility and stability of the resulting closed-loop, for any state in $S_N(\mathcal{X}_s,\mathcal{U})$[2], avoiding the loss of feasibility in presence of changes in the setpoint.

## 3. The proposed MPC

For a fixed control horizon $N \in \mathbb{N}$, we are going to present a novel stable MPC for tracking that provides, for any setpoint in the equilibrium set, $x^* \in \mathcal{X}_s$, the maximal domain of attraction, i.e., the maximal controllable set $S_\infty$. From now on, we simplify the notation by denoting

$$S_i := S_i(\mathcal{X}_s,\mathcal{U}) \quad \text{and} \quad L_i := L_i(\mathcal{X}_s,\mathcal{U}).$$

Let $x \in S_\infty$. The following cost function is proposed

$$V_N(x; \mathbf{u}, \mathbf{x}^a, \mathbf{u}^a, x_s) = \sum_{j=0}^{N-1} \|x_j - x_j^a\|_Q^2 + \|u_j - u_j^a\|_R^2 + V_O(x_s, x^*), \tag{3.1}$$

where, as before, $\mathbf{u} := \{u_0, \cdots, u_{N-1}\}$ represents a sequence of $N$ future control inputs and $x_j$ is the predicted state of the system at time $j$ given by $x_{j+1} = Ax_j + Bu_j$, with $x_0 = x$. The sequence of $N$ state and control auxiliary variables $\mathbf{x}^a := \{x_0^a, \cdots, x_{N-1}^a\}$ and $\mathbf{u}^a := \{u_0^a, \cdots, u_{N-1}^a\}$, have the purpose of accounting for the distance of the states $x_j$ and inputs $u_j$ to some sets that we will define later. Finally, $x_s$ represents an artificial variable in $\mathcal{X}_s$ and $V_O(x_s, x^*) = \|x_s - x^*\|_T^2$.

The controller is derived from the solution of the optimization problem $P_N(x, x^*)$ given by

$$\min_{\mathbf{u},\mathbf{x}^a,\mathbf{u}^a,x_s} V_N(x; \mathbf{u}, \mathbf{x}^a, \mathbf{u}^a, x_s) \tag{3.2}$$
$$\text{s.t.} \quad x_0 = x, \tag{3.3}$$
$$x_{j+1} = Ax_j + Bu_j, \tag{3.4}$$

[2]In the traditional MPC the domain of attraction is $S_N(x^*,\mathcal{U})$, which is significantly smaller than $S_N(\mathcal{X}_s,\mathcal{U})$.

$$x_j \in \mathcal{X}, \quad j \in I_{N-1}, \tag{3.5}$$

$$u_j \in \mathcal{U}, \quad j \in I_{N-1}, \tag{3.6}$$

$$x_j^a \in \Omega_x, \quad j \in I_{N-1}, \tag{3.7}$$

$$u_j^a \in \Psi(\Omega_x), \quad j \in I_{N-1}, \tag{3.8}$$

$$x_N \in \Omega_x, \tag{3.9}$$

$$x_s \in \mathcal{X}_s, \tag{3.10}$$

where $\Omega_x$ is a target set that depends on the initial state $x$, and it is defined as

$$\Omega_x = \begin{cases} S_{kN}, & x \in L_{kN}, \text{ with } k \geq 1 \\ \{x_s\}, & x \in S_N, \end{cases} \tag{3.11}$$

and the set $\Psi(\Omega_x)$ is the corresponding input set to $\Omega_x$. Note that when the state of the system is inside $S_N$ we have that $\Omega_x = \{x_s\}$, therefore the corresponding input set is $\Psi(\Omega_x) = \{u_s\}$, with $u_s$ such that $x_s = Ax_s + Bu_s$.

Considering the receding horizon policy, the control law is given by $\kappa_{MPC}(x) = \mathring{u}_0$, where $\mathring{u}_0$ is the first element of the optimal input sequence $\mathring{\mathbf{u}} = \{\mathring{u}_0, \cdots, \mathring{u}_{N-1}\}$. The optimal cost is defined by

$$\mathring{V}_N(x) := V_N(x; \mathring{\mathbf{u}}, \mathring{\mathbf{x}}^a, \mathring{\mathbf{u}}^a, \mathring{x}_s)$$

where $\mathring{\mathbf{u}}, \mathring{\mathbf{x}}^a, \mathring{\mathbf{u}}^a, \mathring{x}_s$ are the optimal solutions to Problem $P_N(x, x^*)$.

**Remark 3.1.** *The formulation of problem $P_N(x, x^*)$ depends on the computation of the controllable sets $S_{kN}$, for all $k \in \mathbb{N}$. Since $\mathcal{X}$ is a compact set, the sequence $S_k$ converges (up to a given tolerance) in finite steps to the set $S_\infty$. So the number of sets to be computed is finite. On the other hand, these controllable sets depend only on the entire equilibrium set $\mathcal{X}_s$ and not on a specific setpoint $x^*$, i.e. they remain the same regardless of any change on the setpoint. Therefore these sets can be computed off-line, which significantly simplifies the implementation of the optimization problem. This is the main advantage of the proposed approach with respect to the one presented in [19].*

It is easy to see that Problem $P_N(x, x^*)$ presents a dual behaviour just replacing the set $\Omega_x$ on equations (3.7)–(3.9) according to (3.11) and the position of the initial state. We summarized explicitly this dual behaviour in the following properties.

**Property 3.2.** *When the current state $x$ belongs to $S_\infty \setminus S_N$, there exists $k \geq 1$ such that $x \in L_{kN}$, and the artificial variable $x_s$ does not depend on any other optimization variable. Therefore, by optimality, it will be equal to $x^* \in \mathcal{X}_s$ and $\|x_s - x^*\| = 0$. Hence the problem to be solved is equivalent to*

$$\min_{\mathbf{u}} \sum_{j=0}^{N-1} d_Q(x_j, \Omega_x) + d_R(u_j, \Psi(\Omega_x)) \tag{3.12}$$

$$\begin{aligned} \textit{s.t.} \quad & x_0 = x, \\ & x_{j+1} = Ax_j + Bu_j, \quad j \in I_{N-1}, \end{aligned}$$

$$
\begin{aligned}
& x_j \in \mathcal{X}, \quad j \in I_{N-1}, \\
& u_j \in \mathcal{U}, \quad j \in I_{N-1}, \\
& x_N \in \Omega_x,
\end{aligned}
$$

*where $\Omega_x = S_{kN}$, $d_Q(x_j, \Omega_x) := \min\{\|x_j - x_j^a\|_Q^2 : x_j^a \in \Omega_x\}$ is a distance from the predicted state $x_j$ to the target set $\Omega_x$ and $d_R(u_j, \Psi(\Omega_x)) := \min\{\|u_j - u_j^a\|_R^2 : u_j^a \in \Psi(\Omega_x)\}$ is a distance from the control $u_j$ to the set $\Psi(\Omega_x)$. In other words, while the current state does not reach $S_N$, the controller tries to minimize the distance of the predicted trajectory to the next layer.*

**Property 3.3.** *When the controlled system reaches $S_N$, the target set is $\Omega_x = \{x_s\}$. Then by constraint* (3.9) *$x_N = x_s$, by* (3.7) *$x_0^a = x_1^a = \cdots = x_{N-1}^a = x_s$, and by* (3.8) *$u_0^a = u_1^a = \cdots = u_{N-1}^a = u_s$, with $u_s$ such that $x_s = Ax_s + Bu_s$. So the MPCT described in Section 2.1 is recovered.*

### *3.1. Asymptotic stability: preliminaries*

From now on we consider the following assumption.

**Assumption 3.4.** *The set $S_N$ is a contractive CIS.*

**Remark 3.5.** *Note that the above assumption is not so restrictive. For example it is sufficient to have $S_{N-1} \subseteq \operatorname{int}_{S_\infty} S_N$, even when it is not true that $S_{N-1} \subseteq \operatorname{int} S_N$ (see Figure 3). Indeed, if $S_{N-1} \subseteq \operatorname{int}_{S_\infty} S_N$ then $S_N$ is a contractive CIS, and by Lemma 2.6 all $S_{kN}$ are also contractive CIS, for any $k \geq 1$.*

First of all note that the recursive feasibility is an immediate consequence of the nested property of the controllable sets $S_i$. However, the proof of attractivity is more subtle since the optimal cost is not a Lyapunov function in the entire domain of attraction. In order to prove that the real trajectory produced by the proposed strategy reaches the set $S_N$ in a finite number of steps, we need first to suppose the opposite, i.e, we need to proceed by contradiction. The following lemma goes in this direction.

**Lemma 3.6.** *Let $x \in L_{kN}$ for some $k \geq 1$. Let $\{x(i)\}_{i=0}^{\infty}$ be the sequence given by the closed-loop system $x(i+1) = Ax(i) + B\kappa_{MPC}(x(i))$, with $x(0) = x$. If $x(i) \notin S_{kN}$ for all $i \in \mathbb{N}$ then $d_Q(x(i), S_{kN}) \to 0$ when $i \to \infty$.*

*Proof.* Suppose the solution of Problem $P_N(x(i), x^*)$ is given by $\mathring{\mathbf{u}} = \{\mathring{u}_0, \ldots, \mathring{u}_{N-1}\}$, $\mathring{\mathbf{u}}^a = \{\mathring{u}_0^a, \ldots, \mathring{u}_{N-1}^a\}$, $\mathring{\mathbf{x}}^a = \{\mathring{x}_0^a, \ldots, \mathring{x}_{N-1}^a\}$, $\mathring{x}_s = x^*$ and the corresponding optimal state sequence is given by $\mathring{\mathbf{x}} = \{\mathring{x}_0, \ldots, \mathring{x}_N\}$, where $\mathring{x}_0 = x(i)$ and $\mathring{x}_N \in S_{kN}$. The optimal cost is given by

$$
\mathring{V}_N(x(i)) = \sum_{j=0}^{N-1} \|\mathring{x}_j - \mathring{x}_j^a\|_Q^2 + \|\mathring{u}_j - \mathring{u}_j^a\|_R^2.
$$

Since $S_{kN}$ is an invariant set, then there exists $\hat{u} \in \Psi(S_{kN})$ such that $\hat{x} = A\mathring{x}_N + B\hat{u} \in S_{kN}$. Then a feasible solution to problem $P_N(x(i+1), x^*)$ is $\hat{\mathbf{u}} = \{\mathring{u}_1, \ldots, \mathring{u}_{N-1}, \hat{u}\}$, $\hat{\mathbf{u}}^a = \{\overset{o}{u}{}_1^a, \ldots, \overset{o}{u}{}_{N-1}^a, \hat{u}\}$, $\hat{\mathbf{x}}^a = \{\overset{o}{x}{}_1^a, \ldots, \overset{o}{x}{}_{N-1}^a, \overset{o}{x}_N\}$ and $\hat{x}_s = x^*$. The state sequence associated to the feasible input sequence $\hat{\mathbf{u}}$ is given by $\hat{\mathbf{x}} = \{\overset{o}{x}_1, \ldots, \overset{o}{x}_N, \hat{x}\}$. Since $\overset{o}{x}_1 = x(i+1) \notin S_{kN}$ then it is easy to see that $\overset{o}{x}_1 \in L_{kN}$[3]. Therefore the feasible cost corresponding to $\hat{\mathbf{u}}$, $\hat{\mathbf{u}}^a$, $\hat{\mathbf{x}}^a$ and $\hat{x}_s$, is given by

$$\begin{aligned} V_N(x(i+1); \hat{\mathbf{u}}, \hat{\mathbf{x}}^a, \hat{\mathbf{u}}^a, \hat{x}_s) = &\sum_{j=1}^{N-1} \left\{ \|\overset{o}{x}_j - \overset{o}{x}{}_j^a\|_Q^2 + \|\overset{o}{u}_j - \overset{o}{u}{}_j^a\|_R^2 \right\} \\ &+ \underbrace{\|\overset{o}{x}_N - \overset{o}{x}_N\|_Q^2 + \|\hat{u} - \hat{u}\|_R^2}_{=0}, \end{aligned}$$

which means that

$$\begin{aligned} V_N(x(i+1); \hat{\mathbf{u}}, \hat{\mathbf{x}}^a, \hat{\mathbf{u}}^a, \hat{x}_s) - \overset{o}{V}_N(x(i)) &= -\|\overset{o}{x}_0 - \overset{o}{x}{}_0^a\|_Q^2 - \|\overset{o}{u}_0 - \overset{o}{u}{}_0^a\|_R^2 \\ &\leq -\|\overset{o}{x}_0 - \overset{o}{x}{}_0^a\|_Q^2 \\ &= -d_Q(x(i), S_{kN}), \end{aligned} \tag{3.13}$$

where (3.13) is immediate from Property 3.2. Hence the optimal cost $\overset{o}{V}_N(x(i+1))$ satisfies

$$\begin{aligned} \overset{o}{V}_N(x(i+1)) - \overset{o}{V}_N(x(i)) &\leq V_N(x(i+1); \hat{\mathbf{u}}, \hat{\mathbf{x}}^a, \hat{\mathbf{u}}^a, \hat{x}_s) - \overset{o}{V}_N(x(i)) \\ &= -d_Q(x(i), S_{kN}), \end{aligned} \tag{3.14}$$

which implies that $\{\overset{o}{V}_N(x(i))\}_{i=0}^{\infty}$ is a positive decreasing sequence. Thus, $\overset{o}{V}_N(x(\cdot))$ converges to a positive value $\hat{V}$ and, so, $\overset{o}{V}_N(x(i+1)) - \overset{o}{V}_N(x(i)) \to 0$ when $i \to \infty$. Therefore, by (3.14), $d_Q(x(i), S_{kN}) \to 0$ when $i \to \infty$.

### *3.2. Asymptotic stability: main theorem*

**Lemma 3.7 (Stepping through the layers).** *Let $x \in L_{kN}$ for $k \geq 1$. System* (2.1) *controlled by the implicit law $\kappa_{MPC}(\cdot)$, provided by problem $P_N(x, x^*)$, reaches the next layer $L_{(k-1)N}$.*

*Proof.* We proceed by contradiction. Suppose that the sequence $\{x(i)\}_{i=0}^{\infty}$ given by the closed-loop system $x(i+1) = Ax(i) + B\kappa_{MPC}(x(i))$, with $x(0) = x$, does not reach the next layer $L_{(k-1)N}$. Then $x(i) \notin S_{kN}$, for any $i \in \mathbb{N}$. Therefore, by Lemma 3.6, $d_Q(x(i), S_{kN}) \to 0$, when $i \to \infty$.

Since by Lemma 2.6 $S_{kN}$ is a compact and convex contractive CIS, then by Lemma 2.5 $S_{kN} \subset \text{int}_{S_\infty} S_1(S_{kN}, \Psi(S_{kN}))$. Hence there exists $i_0 \in \mathbb{N}$ such that $x_0 := x(i_0) \in S_1(S_{kN}, \Psi(S_{kN}))$. Therefore there exists $u_0 \in \Psi(S_{kN})$ such that $x_1 = Ax_0 + Bu_0 \in S_{kN}$. From the contractive invariance of $S_{kN}$ there exist $u_1, \ldots, u_{N-1} \in \Psi(S_{kN})$ such that $x_{j+1} = Ax_j + Bu_j \in \text{int}_{S_\infty} S_{kN} \subset S_{kN}$, for $j = 1, \ldots, N-1$.

[3] Note that $\overset{o}{x}_1 \notin S_{(k+2)N}$, otherwise $\overset{o}{x}_N$ it could not reachs the set $S_{kN}$ in $N-1$ steps.

Since we are under the assumptions of Property 3.2, the cost function for this control sequence is

$$\begin{aligned} & d_Q(x_0, S_{kN}) + \underbrace{d_R(u_0, \Psi(S_{kN}))}_{=0} \\ & + \underbrace{\sum_{j=1}^{N-1} d_Q(x_j, S_{kN}) + d_R(u_j, \Psi(S_{kN}))}_{=0} = d_Q(x_0, S_{kN}), \end{aligned}$$

while any control action that leaves $x_1$ outside $S_{kN}$ produces a cost grater than $d_Q(x_0, S_{kN})$. Thus, $x(i_0 + 1) = Ax(i_0) + B\kappa_{MPC}(x(i_0)) \in S_{kN}$, which contradicts the fact that $x(i) \notin S_{kN}$, for all $i \in \mathbb{N}$.

Now we have all the ingredients necessary to present and prove the main result of this work.

**Theorem 3.8 (Attractivity of $x^*$ in $S_\infty$).** *Let $x \in S_\infty$. Let $\{x(i)\}_{i=0}^{\infty}$ be the sequence given by the closed-loop system $x(i+1) = Ax(i) + B\kappa_{MPC}(x(i))$, with $x(0) = x$. Then $d_Q(x(i), x^*) \to 0$, when $i \to \infty$.*

*Proof.* Since $x \in S_\infty$ then, by (2.2), $x \in S_N$ or there exists $k_0 \geq 1$ such that $x \in L_{k_0 N}$. In the first case ($x \in S_N$), problem $P_N(x, x^*)$ becomes the tracking problem (see Property 3.3). So, the recursive feasibility and the attractivity of $\{x^*\}$ can be proved by means of the same arguments as in [21]. In the second case, the recursive feasibility can be easily obtained by induction, noticing that any state in a $kN$-Layer belongs to the set $S_{(k+1)N} = S_N(S_{kN})$, and so, there exists a feasible trajectory which drives the closed-loop to $S_{kN}$ in $N$ steps. On the other hand, since $x \in L_{k_0 N}$ we can apply recursively Lemma 3.7 until the current state reaches the set $S_N$, which means that we are again under the conditions of the first case and, therefore, the attractivity is proved.

**Corollary 3.9 (Asymptotic stability).** *The setpoint $\{x^*\}$ is asymptotically stable for the closed-loop system controlled by $\kappa_{MPC}(\cdot)$, for all $x \in S_\infty$.*

*Proof.* Since $x^* \in \mathcal{X}_s \subseteq S_N$, our strategy inherits the local stability for any $x \in S_N$ from the MPCT, using the same Lyapunov function as in [21]. Then, the asymptotic stability is a straightforward consequence of Theorem 3.8.

## 4. Illustrative Example

In this section some simulations results will be presented to evaluate the proposed control strategy. To this end, let us consider a constrained sampled double integrator. Although simple, this second order system, which models the dynamic of a mass (position/velocity) in a one dimensional space, is a classic control benchmark, since it allows one to test and graphically show the properties of a certain control strategy. In the context of the present work, it allows us to observe some details of the MPC controller presented in this note, such as the importance of considering the concept of interior relative to the set $S_\infty$ addressed in Remark 2.7.

The discrete-time model of the double integrator here considered, is given by

$$x(i+1) = \begin{bmatrix} 1 & 1 \\ 0 & 1 \end{bmatrix} x(i) + \begin{bmatrix} 0 & 0.5 \\ 1 & 0.5 \end{bmatrix} u(i), \tag{4.1}$$

with the constraints $\mathcal{X} = \{x \in \mathbb{R}^2 : |x_1| \leq 5, |x_2| \leq 1\}$ and $\mathcal{U} = \{u \in \mathbb{R}^2 : \|u\|_\infty \leq 0.05\}$.

### *4.1. Domain of attraction and layers*

The main advantage of the proposed strategy is that the largest domain of attraction can be obtained independently of the prediction horizon chosen for the design of the MPC. That is, even taking the smallest admissible horizon, the maximal domain of attraction will be reached.

In Figure 2 the domain of attraction of the proposed controller, $S_\infty$, is compared with the one provided by MPCT, $S_N(\Omega_t, \mathcal{U})$, presented in [21], for a prediction horizon $N = 3$. As it can be seen, the domain of attraction of the proposed MPC is significantly larger than the one of the MPCT for the selected horizon.

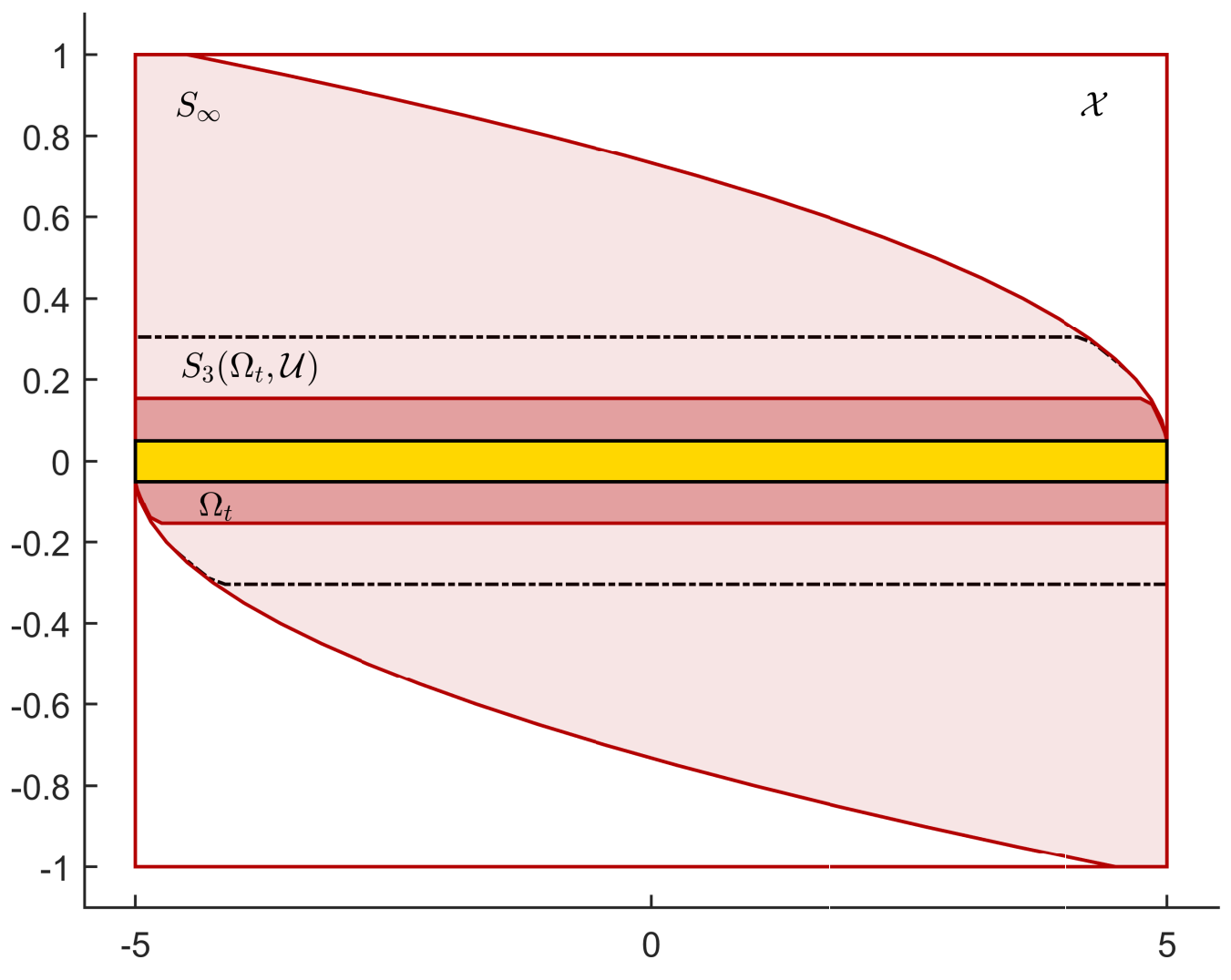

Figure 2: $S_3(\Omega_t,\mathcal{U})$: Domain of attraction of the MPCT, with prediction horizon $N = 3$. $S_\infty$: Domain of attraction of the proposed MPC, with the same prediction horizon.

On the other hand, to use the proposed strategy we need to compute the sets that constitutes the layers that are used as terminal constraint in (3.10).

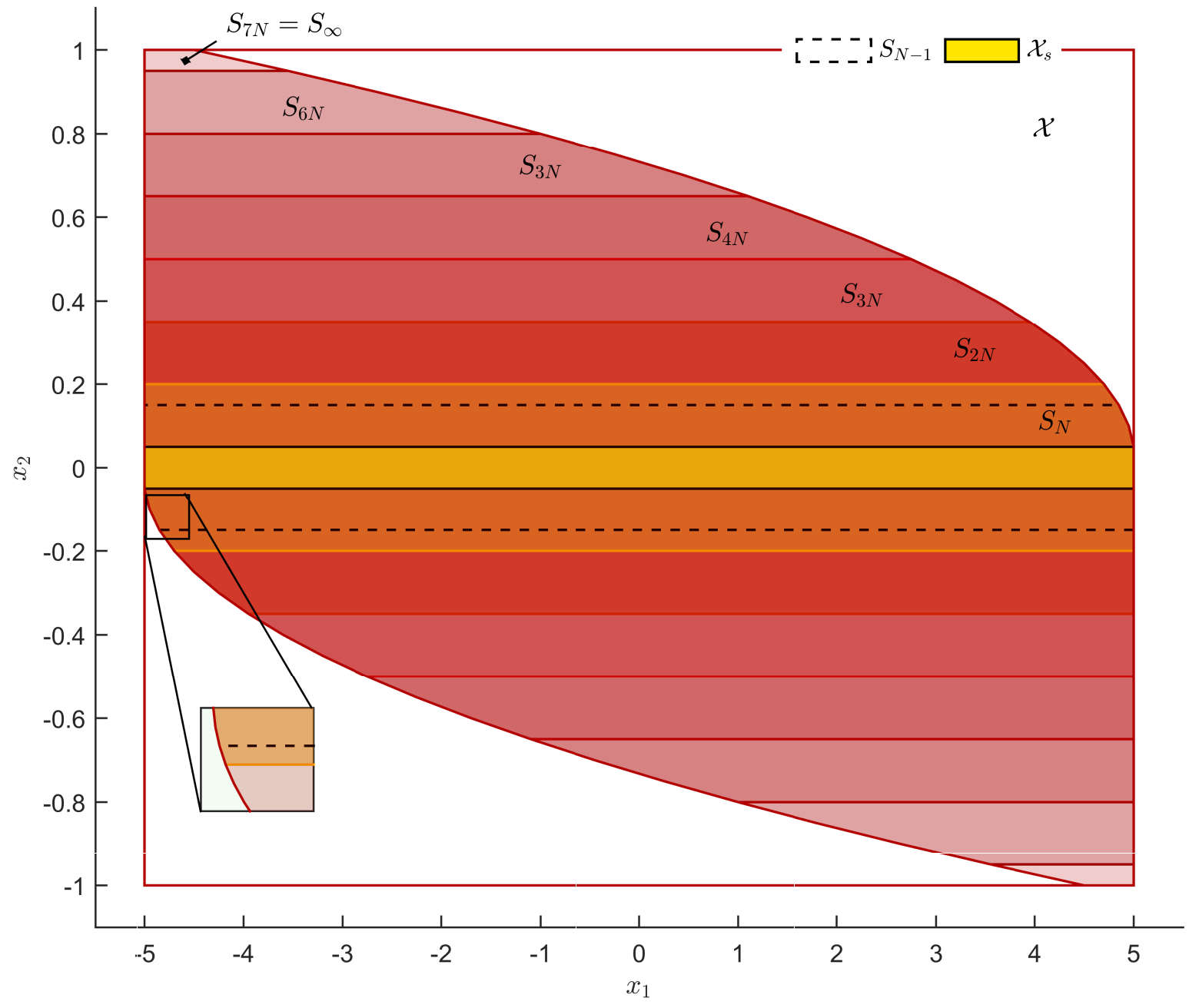

Figure 3: $S_{kN}$ for $k = 1, \ldots, 7$ and $N = 3$. Zoom window shows that the set $S_{N-1} \subset \text{int}_{S_\infty} S_N$.

Figure 3 shows the equilibrium set $\mathcal{X}_s$, the controllable set $S_{N-1}$, and a sequence of sets $S_{kN}$, $k \in \mathbb{N}$, with a prediction horizon $N = 3$. Observe that $S_{N-1} \subset \text{int}_{S_\infty} S_N$, which implies that $S_N$ is a contractive CIS (see Remark 3.5). So, the propose MPC will be tested under the Assumption 3.4. Note also that the maximal domain of attraction of system (4.1) is reached for $k = 7$, i.e $S_\infty = S_{7N}$.

**Remark 4.1.** *It should be noted that the calculation of the layers sets is obtained only once and off-line, that is, their computation does not interfere with the performance of the controller. It is clear that the computational complexity of the sets involved in the proposal increases with the order of the system. If the dimension of the system is small, the computation of such sets as polyhedron can be easily managed with tool such as the MPT 3.0 toolbox [11]. On the other hand, for large scale systems the zonotopes representation[4] –which can be managed with toolboxes like CORA– is more appropriate [17, 26].*

---

[4]A zonotope of order $m$ is the set given by $p \oplus H\mathbf{B}^m = \{p + H\varepsilon : \varepsilon \in \mathbf{B}^m\}$, where the vector $p \in \mathbb{R}^n$ represents the center of the zonotope, $H \in \mathbb{R}^{n \times m}$ represents the matrix of generators, and $\mathbf{B}^m$ is a unitary box centered in the origin. Note that a zonotope is the Minkowski sum of the segments defined by the columns of matrix H.

Table 1: State and Input Constraints

| Variables | Constraints | Unit |
|---|---|---|
| $(x, y)$ | $[-6, 6]$ | $m$ |
| $z$ | $[1, 7]$ | $m$ |
| $(\dot{x}, \dot{y}, \dot{z})$ | $[-5, 5]$ | $m/s$ |
| $(\phi, \theta, \psi)$ | $[-\frac{\pi}{4}, \frac{\pi}{4}]$ | $rad$ |
| $(\dot{\phi}, \dot{\theta}, \dot{\psi})$ | $[-3, 3]$ | $rad/s$ |
| $u_1$ | $[0, g + 2.38]$ | $m/s^2$ |
| $(u_2, u_3, u_4)$ | $[-0.5, 0.5]$ | $rad/s^2$ |

*4.2. Domain of attraction and layers for higher order systems*

To show the applicability of the proposal to the case of higher order systems, we choose the autonomous quadrotor benchmark proposed in [14, 13], which consists of a 12-dimensional system. The state vector is given by

$$x = \begin{bmatrix} x & y & z & \dot{x} & \dot{y} & \dot{z} & \phi & \theta & \psi & \dot{\phi} & \dot{\theta} & \dot{\psi} \end{bmatrix}' \in \mathbb{R}^{12}$$

where $(x, y, z)$ and $(\dot{x}, \dot{y}, \dot{z})$ represent translational positions (in $[m]$) and velocities (in $[m/s]$), while $(\phi, \theta, \psi)$ and $(\dot{\phi}, \dot{\theta}, \dot{\psi})$ represent Eulerian angles (in $[rad]$) and angular velocities (in $[rad/s]$). The control input is given by the vector

$$u = \begin{bmatrix} u_1 & u_2 & u_3 & u_4 \end{bmatrix}' \in \mathbb{R}^4$$

which gathers the total normalized thrust (in $[m/s^2]$) and the angular acceleration $(\ddot{\phi}, \ddot{\theta}, \ddot{\psi})$ (in $[rad/s^2]$).

The nonlinear continuous time equations of motion that model the dynamic of the system can be found in [14] and are given by:

$$\ddot{x} = -u_1 \cos(\phi) \sin(\theta) \quad (4.2a)$$

$$\ddot{y} = u_1 \sin(\phi) \quad (4.2b)$$

$$\ddot{z} = -g + u_1 \cos(\phi) \cos(\theta) \quad (4.2c)$$

$$\ddot{\phi} = u_2 \quad (4.2d)$$

$$\ddot{\theta} = u_3 \quad (4.2e)$$

$$\ddot{\psi} = u_4 \quad (4.2f)$$

being $g = 9.81$ the acceleration of gravity. A discrete-time LTI model can be obtained by linearizing model (4.2) about the hover condition $\phi = 0$, $\theta = 0$ and $u_1 = g$, and discretizing with sample time $T_s = 0.2\ [sec]$. The state and input constraints are shown in Table 4.2.

In order to compute the controllable set $S_N$, and the sequence of sets $S_{kN}$, $k \in \mathbb{N}$, we first identified the minimum control horizon, say $N^*$, that ensures controllability of the system. The value that we obtained is $N^* = 4$ and it represents the minimum $N$ such that the controllability matrix

$$\left[A^{N-1}B, A^{N-2}B, ..., AB, B\right]$$

is full rank. For simulation we choose a control horizon $N = 5$. With this choice, we can reach the maximal domain of attraction of the MPCT controller with $k = 7$, which is in accordance with the fact that the maximal domain of attraction $S_\infty$ of standard MPCT [20, 8] is reached with a control horizon of $N = 34$.

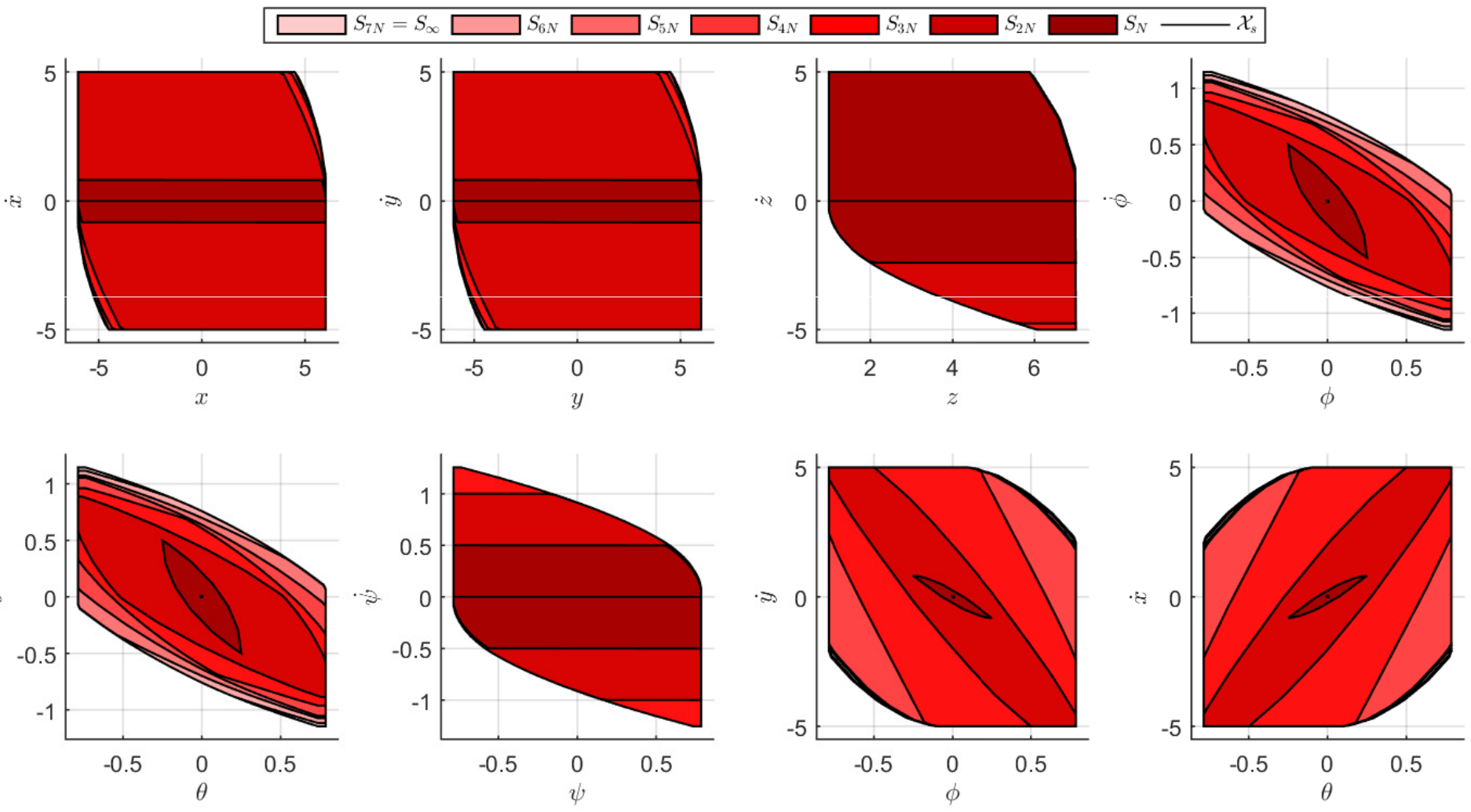

Figure 4: Sequence of controllable sets $S_{kN}$ for $k = 1, \ldots, 7$ and $N = 5$ (projection onto linear and angular position/velocity dimensions).

Figure 4 shows the projections of the obtained sets onto (linear and angular) position/velocity dimensions. The projections of the equilibrium set $\mathcal{X}_s$ are also plotted (in black). It is worth remarking that all sets involved in this simulation are computed as constrained zonotopes, by means of the MATLAB Toolbox CORA [2, 1]. This way we can compute the sequence of sets $S_{kN}$, $k \in [1, 7]$ in 6.15 seconds. The maximal domain of attraction of standard MPCT has been computed in 21.26 seconds. All computations were run in Matlab R2020a on a MacBookPro running macOS 10.15.5 with a Quad-Core Intel Core i5 processor at 2.3 Ghz and 8 GB of RAM. This example shows that the use of constrained zonotopes [26] allows one to compute the sets necessary for the applicability of the proposed approach, even in case of a higher dimensional systems.

### *4.3. Performance of the proposed controller*

Let us consider now the double integrator (4.1). The control objective will be to bring the system to two different setpoints. To test the dynamic performance of the closed-loop system controlled by the proposed MPC, it is considered a starting point in the farthest layer from the equilibrium set, $x_0 = (-4.9; 0.96)$. Besides, a setpoint change has been considered: the setpoint is first $x_i^* = (-4; 0)$, but before the closed-loop system reaches it, the setpoint is switched to $x_f^* = (3.5; 0)$, at time $k = 70$. The state space evolution in Figure 5 clearly shows the capability of the proposed controller to drive the closed-loop system toward the desired setpoint, without loss of feasibility.

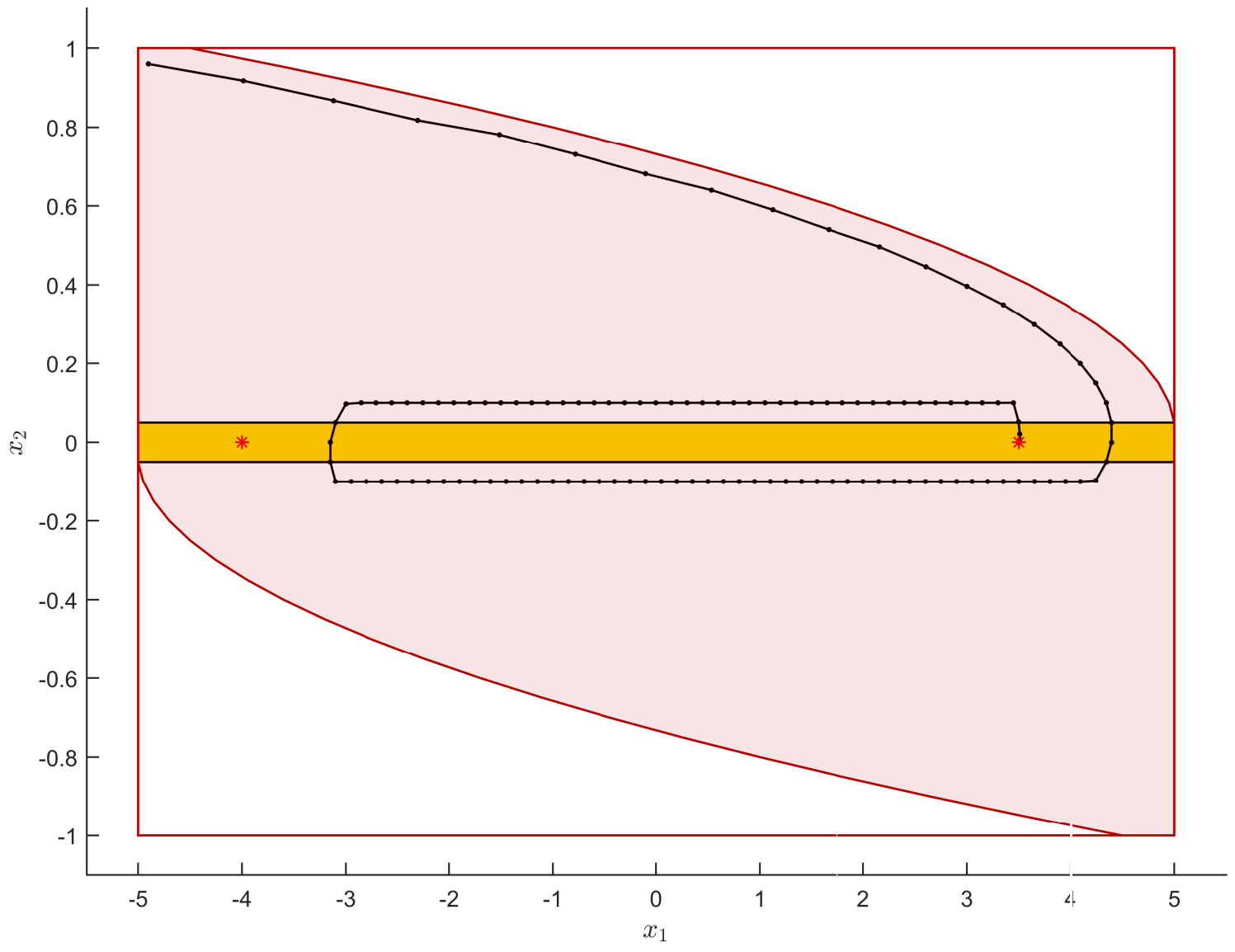

Figure 5: Closed-loop system evolution starting from $x_0 = (-4.9; 0.96)$. For time $k \leq 70$ the setpoint is $x_i^* = (-4; 0)$ and it is $x_f^* = (3.5; 0)$ for $k > 70$.

In what follows, the performance of the proposed MPC is compared with two other strategies, who also seek to solve this problem. The first one is the MPC presented in [19], which also exhibits the maximal domain of attraction that the system allows for any prediction horizon. The disadvantage of this strategy is that it does not contemplate possible setpoint changes. This means that given changes in the objective to be achieved, loss of feasibility may occur. In this context, the advantage of the proposed controller is notable.

A second comparison will be performed with the MPCT proposed in [21], which solves the problem of loss of feasibility under changes in the setpoint and enlarges the domain of attraction of the controller.

In order to properly asses the performances we proposed the following index

$$\Phi = \frac{1}{T_{sim}} \sum_{k=1}^{T_{sim}} \|x(k) - x^*\|_\infty + \|u(k) - u^*\|_\infty, \tag{4.3}$$

where $T_{sim}$ represents the total simulation time. Index $\Phi$ penalizes the distance - given by the infinite norm - between the states and inputs of the closed-loop system with respect to the given setpoint.

The strategy which we will compare the performance of the proposed controller with, is the MPC for tracking with a terminal cost function and terminal inequality constraint, proposed in [21]. This choice is made in order to compare the proposed MPC with the MPCT formulation that provides the best performance. The terminal cost function of the MPCT is taken as $V_f(x - x_a) = \|x - x_a\|_P^2$, with P solution of the Lyapunov equation. The terminal constraint is given by $(x(N), x_a, u_a) \in \Omega_t^a$, where $\Omega_t^a$ is the so-called invariant set for tracking [20]. The local (terminal) controller has been chosen as the Linear Quadratic Regulator (LQR) with $Q = I_n$ and $R = 2I_m$, and it is given by

$$K_{LQR} = \begin{bmatrix} 0.0509 & -0.3910 \\ -0.4335 & -0.7736 \end{bmatrix}. \tag{4.4}$$

Let $\Omega_t$ be the projection of $\Omega_t^a$ onto $\mathcal{X}$, then the domain of attraction is given by the set of states that can be admissible steered to the set $\Omega_t$ in $N$ steps, i.e. $S_N(\Omega_t, \mathcal{U})$.

Figure 2 compares the domain of attraction of the MPCT, $S_N(\Omega_t, \mathcal{U})$, with the domain of attraction of the proposed MPC, $S_\infty$, with prediction horizon $N = 3$ in both case. As it can be seen, the domain of attraction for the proposed MPC is significantly larger than the MPCT for the selected horizon. Even more, to reach the maximal domain of attraction with the MPCT, we would need a prediction horizon of $N = 18$, i.e. $S_\infty = S_{18}(\Omega_t, \mathcal{U})$, which would produce a non negligible increase in the computational cost.

The numerical comparison between the performance of both controllers was made by using Index (4.3). To carry out this comparison, several initial random points were taken within the set $S_\infty$. Each initial point is steered to the given setpoint by both controllers. The proposed MPC design in this experiment used a prediction horizon $N = 3$. The MPCT is not able to control every point of $S_\infty$ with $N = 3$, so it is designed with horizon $N = 18$. The average values of the Index is shown in Table 2.

As expected, the performance of the proposed controller is not better than the one of the MPCT. In fact, the better performance of the MPCT is justified by the larger prediction horizon ($N = 18$). Anyway, it should be noted that the performance difference is not significant (2.12%), and seems to be a reasonable price to pay to obtain a meaningful prediction horizon reduction ($N = 3$). In particular, Figures 6 and 7 show, respectively, the space and time system evolution for the proposed MPC and MPCT. It can be seen that both strategies start from point $x_0 = (-4.9; 0.96)$ and converge to $x^* = (-4; 0)$, even if, as expected, the closed-loop trajectories that they produce are completely different.

In the next simulations, the strategies will be compared taking the same prediction horizon $N = 3$ and using terminal equality constraint in the MPCT formulation. Since the initial condition $x_0 = (-4.9; 0.96)$, is not inside the domain of attraction of the MPCT, $x_0 \notin S_3$, the performance of the two controllers will be compared only considering the system evolution inside set $S_3$. The result is shown in the Table 2. Furthermore, Figures 8 and 9 show, respectively, space and time system evolution for the proposed MPC and MPCT. It can be seen that the proposed MPC starts from point $x_0 = (-4.9; 0.96)$. On the other hand, the MPCT controller has $x_0^2 = (4.0987; 0.1999)$ as its initial condition, since $x_0$ does not belong to its domain of attraction. As expected, inside set $S_3$, the proposed controller presents same performance than the MPCT. The difference is that the MPCT controller cannot be feasible from $x_0$ as initial condition, since its domain of attraction is significantly smaller than the one provided by the controller proposed in this work.

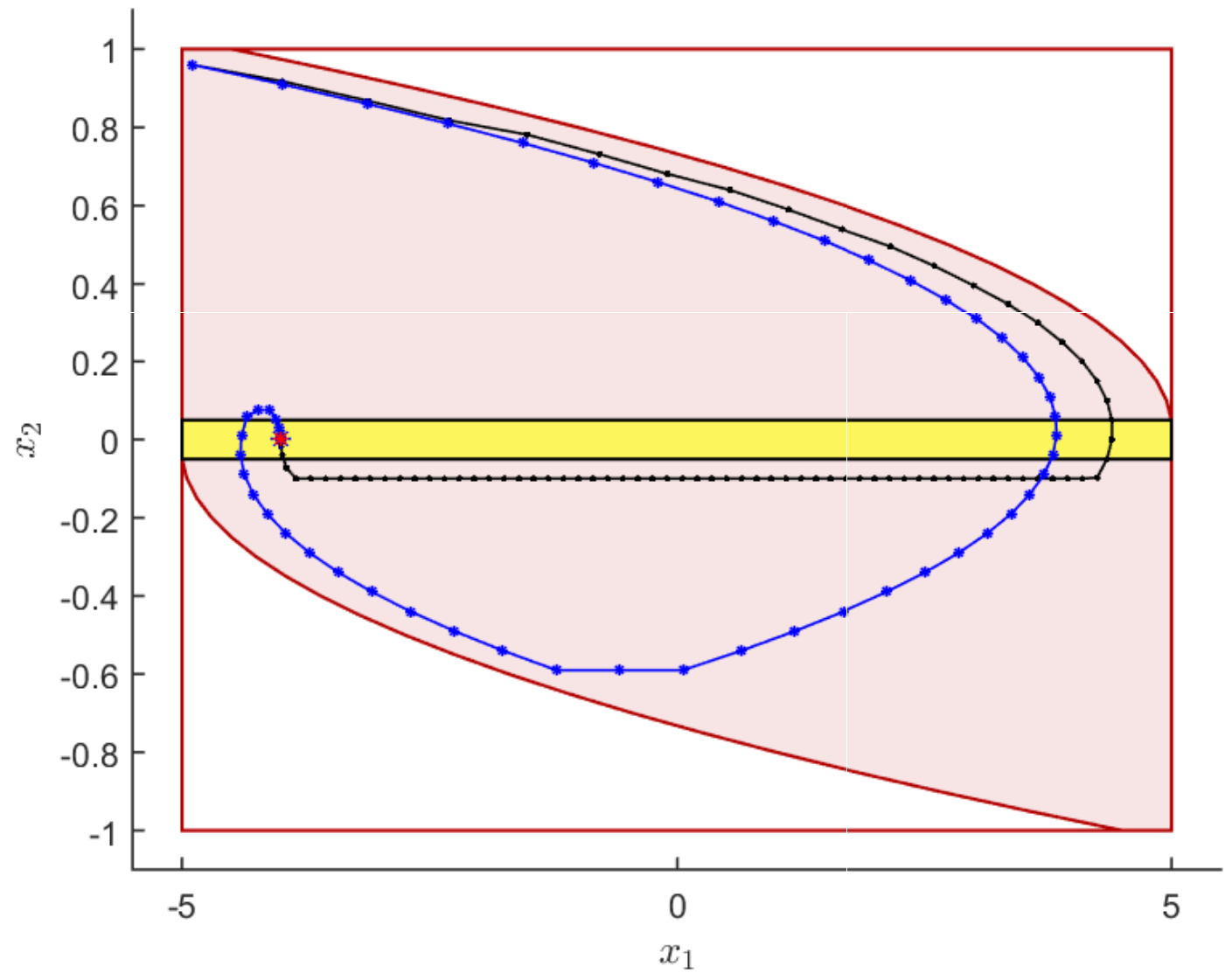

Figure 6: Closed-loop system evolution starting from $x_0 = (-4.9; 0.96)$. System evolution using the proposed MPC with $N = 3$ in black line, evolution using MPCT with $N = 18$ in blue line

| | Average of $\Phi$ (different horizon) | Average of $\Phi$ (same horizon) |
|---|---|---|
| Proposed MPC | 2.0480 | 3.3691 |
| MPCT [21] | 2.0053 | 3.3691 |

Table 2: Performance of the proposed MPC and the MPCT presented in [21]. The second column shows the results considering the proposed MPC strategy with an predictive horizon $N = 3$ and the MPCT strategy with $N = 18$. The third column shows the results considering the same horizon $N = 3$ for each strategy.

## 5. Conclusions

A novel set-based MPC for tracking was presented, which achieves the maximal domain of attraction that the constrained system under control permits. The formulation consider a fixed (arbitrary small) prediction/control horizon and, opposite to other existing strategies, have proved to be recursively feasible and asymptotically stable under any possible change of the setpoint. Furthermore, it preserves the optimizing behavior (i.e., it does not only pass from one state space region to the next, but also minimizes a cost function in the path) for every initial condition in the domain of attraction.

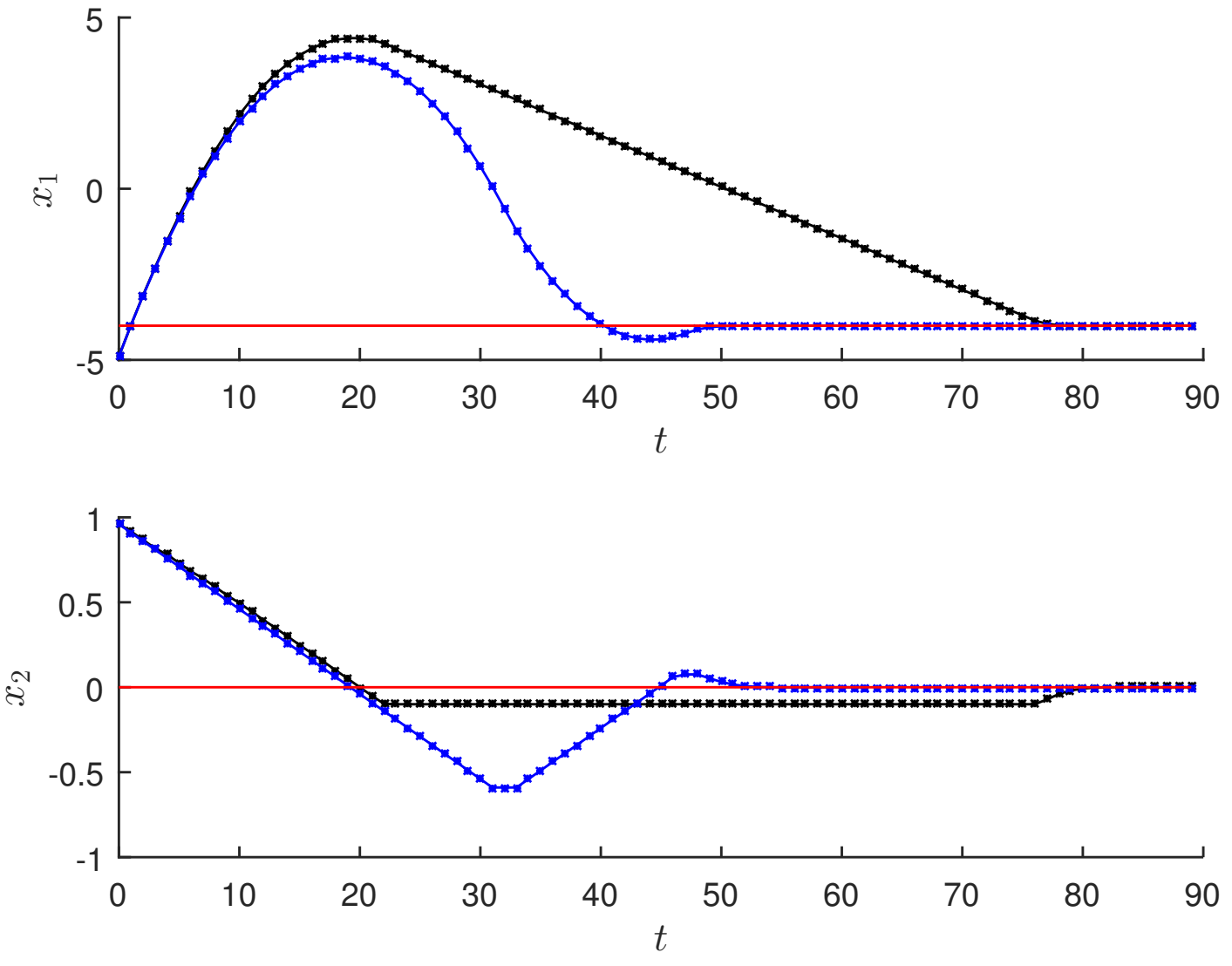

Figure 7: Time system evolution starting from $x_0 = (-4.9; 0.96)$. System evolution using the proposed MPC with $N = 3$ in black line, evolution using MPCT with $N = 18$ in blue line and optimal reference in red line.

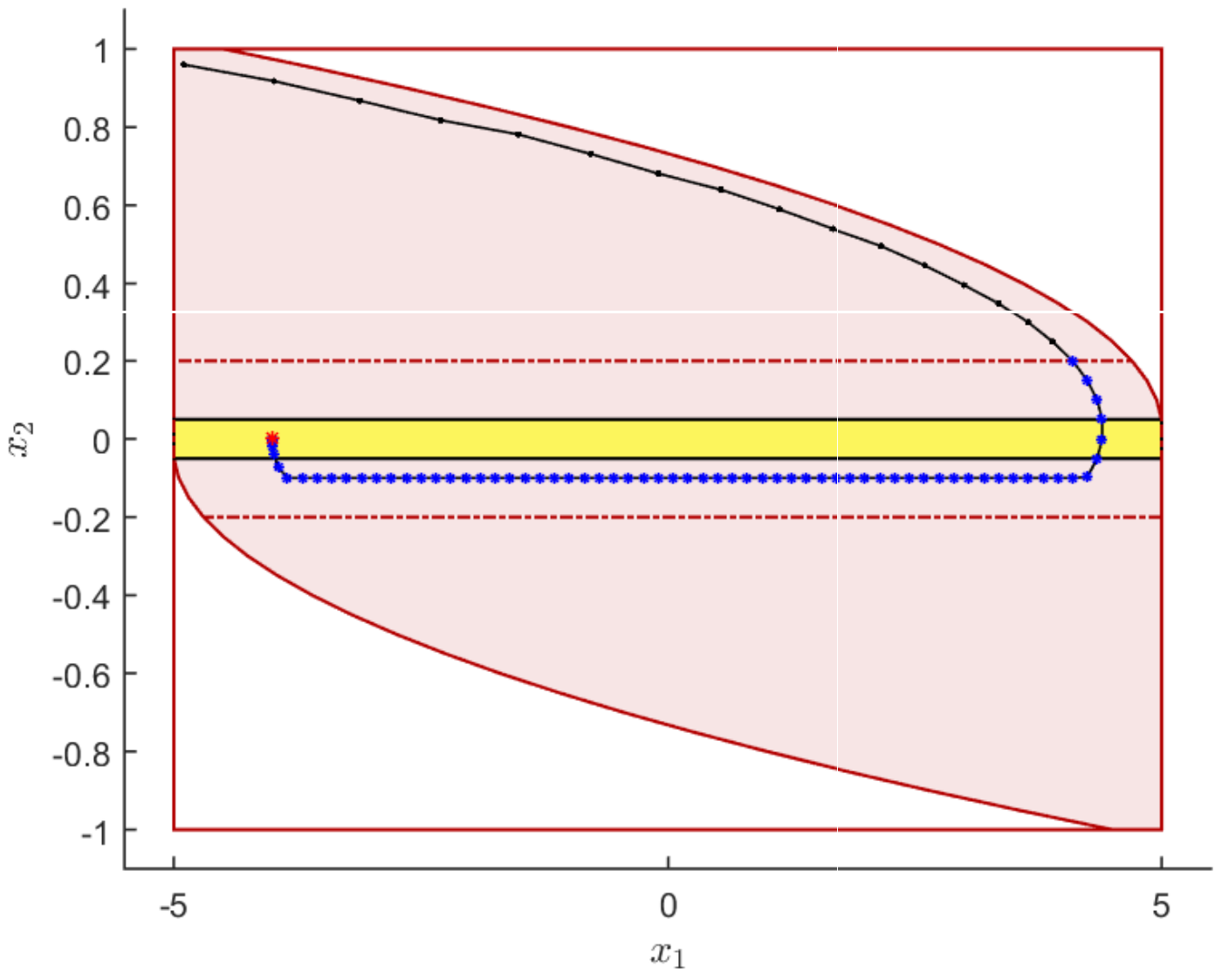

Figure 8: Closed-loop system evolution starting from $x_0 = (-4.9; 0.96)$. System evolution using the proposed MPC with $N = 3$ in black line, evolution using MPCT with $N = 3$ in blue line.

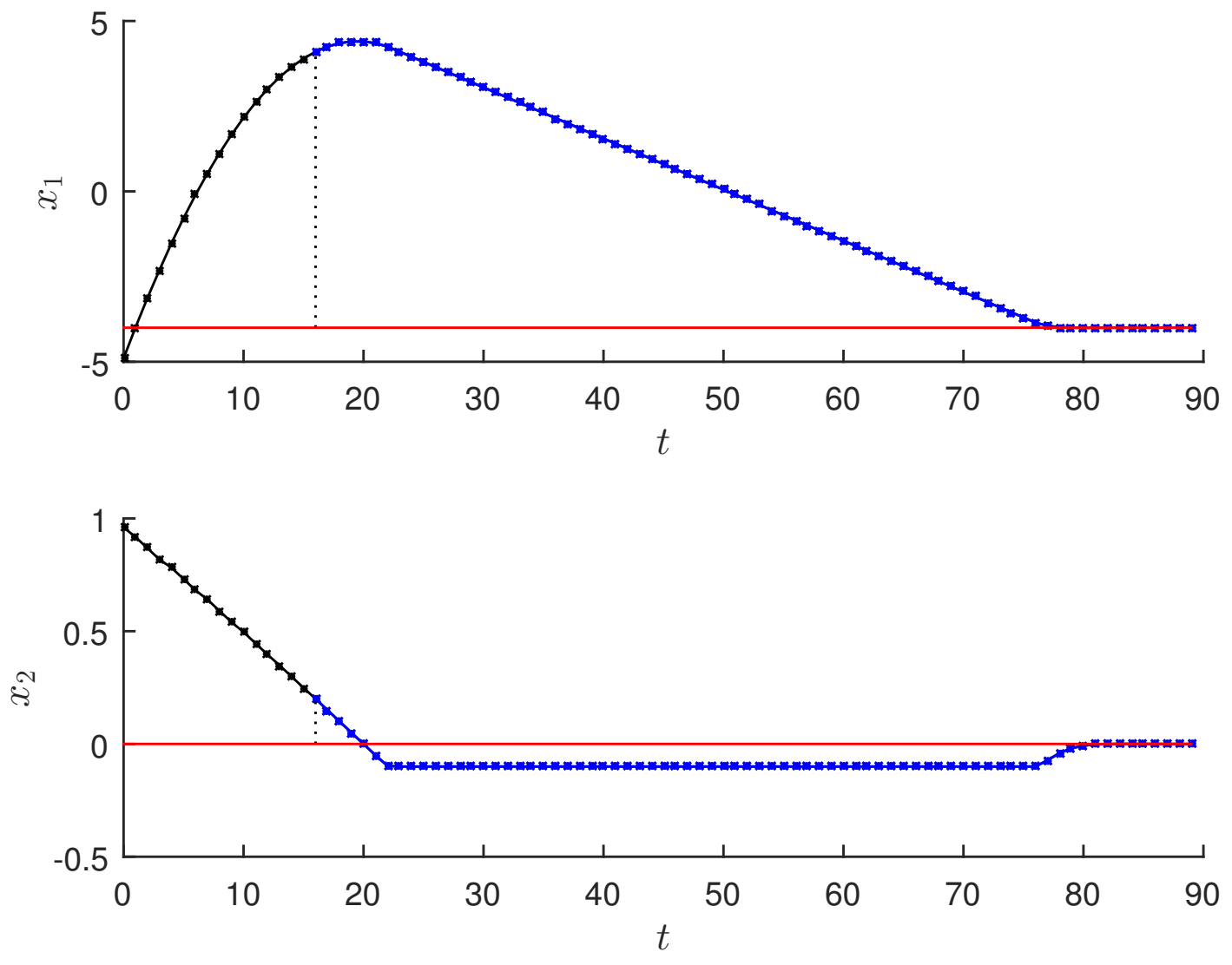

Figure 9: Time system evolution starting from $x_0 = (-4.9; 0.96)$. System evolution using the proposed MPC with $N = 3$ in black line, evolution using MPCT with $N = 3$ in blue line and optimal reference in red line.

These benefits are achieved by solving a rather simple on-line, set-based, optimization problem, which depends on the off-line computation of a sequence of fixed controllable sets (in contrast to what is made, for instance, in [19], where the sets depend on the setpoint). The resulting controller has been successfully compared with other methods, by means of several simulation examples. Future works include more challenging application examples and a detailed robust analysis/extension.